# Extraordinary oxidation behavior of W-Zr thin-film metallic glasses: A route for tailoring functional properties of W-Zr-O films

Petr Zeman[*], Michaela Červená, Jiří Houška, Stanislav Haviar, Jiří Rezek, Šárka Zuzjaková

*Department of Physics and NTIS, European Centre of Excellence, University of West Bohemia, Univerzitní 8, 301 00 Plzeň, Czech Republic*

**Abstract**

The oxidation behavior of W-Zr thin-film metallic glasses (TFMGs) with 32, 48 and 61 at.% Zr, prepared by dc magnetron co-sputtering, was comprehensively studied after annealing in synthetic air. The study focuses on the effect of the annealing temperature (up to 600°C) on the oxidation process, oxygen saturation, structure evolution, and their subsequent impact on electrical, optical and mechanical properties. The findings reveal that controlled oxidation transforms W-Zr TFMGs into amorphous ceramic W-Zr-O films with substoichiometric compositions. This is a consequence of an oxidation process that does not proceed through the formation of a stoichiometric oxide layer on the surface of W-Zr TFMGs, acting as a diffusion barrier against fast oxidation, but leads to a gradual incorporation of oxygen across the film volume due to thermodynamics factors. Higher Zr content accelerates the oxygen incorporation and its depth uniformity in the films. As a result, the mechanical properties are significantly enhanced achieving hardness values of up to 17.5 GPa at approximately 50% oxygen saturation. Simultaneously, the electrical and optical properties are finely tuned with the resistivity and the extinction coefficient (measured at 550 nm) ranging from 1.7 to 95.7×10$^{-4}$ Ω·cm and 0.28 to 1.06, respectively.

\* Corresponding author.

E-mail address: zemanp@kfy.zcu.cz (P. Zeman)

# 1. Introduction

Thin-film metallic glasses (TFMGs) represent a unique class of metallic amorphous films that combine the exceptional properties of bulk metallic glasses (BMGs) with the advantages of thin-film fabrication [1–3]. Using magnetron sputter deposition, a highly non-equilibrium process, TFMGs can be synthesized with as few as two elements and their composition can be easily tuned across a wide range [4–8]. Their amorphous structure with the structural and compositional homogeneity, coupled with properties such as high strength, superior elastic limits, enhanced plasticity, corrosion resistance and thermal stability, makes TFMGs highly versatile. They are particularly attractive for advanced applications, including protective coatings, flexible electronics, biomedical devices, and structural components in state-of-the-art engineering systems.

As oxidation is a critical factor that directly impacts the long-term performance and durability of materials, understanding of the oxidation behavior of TFMGs is essential. Although studies on this topic are relatively scarce, the available research presents promising findings that position TFMGs as an attractive class of materials also in this context. Notably, a few comparative studies on the oxidation behavior of TFMGs and their crystalline counterparts in two binary model systems, Cu-Zr and Al-Zr, have been published [9–11].

These studies revealed that the oxidation kinetics of TFMGs is significantly slower compared to that of crystalline alloys with the same composition. The absence of long-range atomic ordering in TFMGs eliminates grain boundaries and crystal defects as common low-density channels or sites for the fast inward diffusion of oxygen and outward diffusion of alloy constituents. Consequently, the onset of oxidation in TFMGs occurs at higher temperatures [9], and the oxide scale that forms on their surface is amorphous [10–13]. However, the oxidation behavior of the Cu-Zr and Al-Zr systems exhibits also notable differences. In the Cu-Zr system, Zr has a much higher affinity for oxygen than Cu and forms zirconia with a more negative enthalpy of formation compared to copper oxides. This results in an oxide scale comprising amorphous $ZrO_2$ with a Cu-rich, Zr-depleted transition zone beneath it [10]. In contrast, the Al-Zr system is characterized by similar oxygen affinities for Al and Zr and thus comparable enthalpies of formation for their respective oxides. As a result, the oxide scale formed on Al-Zr TFMGs, independently of the alloy composition, is an amorphous metastable $(Al_{0.33}Zr_{0.67})O_{1.83}$ phase with a stoichiometry consistent with $Al_2O_3$ and $ZrO_2$ [12].

The oxidation kinetics of Zr-based TFMGs and BMGs typically obey the parabolic rate law below their crystallization temperature. This behavior is characteristic of a diffusion-controlled process, where the transport of ionic species occurs through an oxide scale, which acts as a barrier layer between the surface and the underlying alloy [10–19]. However, our recent investigations [20] have revealed that W-Zr TFMGs prepared by magnetron sputter deposition deviate from this standard oxidation behavior, suggesting different oxidation mechanisms for these materials. While crystalline W-Zr films with up to 24 at.% Zr oxidize very slowly due to the formation of the oxide scale, W-Zr TFMGs with the Zr content between 33 and 83 at.% Zr consume oxygen much faster due to the absence of a protective oxide surface layer. Interestingly, this oxidation behavior does not lead to their catastrophic degradation but instead results in the transformation of the W-Zr TFMGs to compact, homogeneously oxidized substoichiometric W-Zr-O films with an amorphous structure and enhanced properties when annealed in synthetic air to 600°C.

The present work aims at extending these remarkable findings through a systematic investigation of the effect of controlled post-deposition oxidation of W-Zr TFMGs with three different compositions at various annealing temperatures. The goal is to establish a convenient strategy for preparing amorphous W-Zr-O films with tailored optical, electrical and mechanical properties, and thereby presenting their potential for future applications.

## 2. Experimental

*2.1. Deposition*

Three binary W-Zr TFMGs were deposited onto unheated, unbiased Si(100) or pre-oxidized Si(100) substrates using dc magnetron co-sputtering of W (99.95%) and Zr (99.7%) targets. The substrates were rotated at 40 rpm above the targets to ensure the thickness and composition uniformity. The composition of the films (in at.%) was $W_{68}Zr_{32}$, $W_{52}Zr_{48}$ and $W_{39}Zr_{61}$, and are hereafter referred to as W-rich, W/Zr balanced and Zr-rich, respectively. More detailed information on other deposition parameters can be found in our previous papers [8,20].

*2.2. Post-deposition oxidation*

To investigate the oxidation behavior, the as-deposited films were annealed in synthetic flowing air (1 l/h) in either a high-resolution Setaram TAG 2400 thermogravimeter to 600°C or in a high-temperature Clasic Vac 1800 furnace to various temperatures ranging from 300°C to 600°C. In both setups, the heating rate was 10°C/min and the films were immediately cooled down to room temperature after reaching the final preset temperatures.

*2.3. Characterization*

The elemental composition of the films on the Si(100) substrates was measured by scanning electron microscopy (SEM) using a Hitachi SU-70 microscope operated at a primary electron energy of 10 keV using a Thermo Scientific UltraDry energy dispersive spectrometer (EDS). In the case of the measurement of cross-sectional EDS profiles, the primary electron energy was 5 keV. The film samples were encapsulated in resin, mechanically cut and polished, and finally polished by the argon ion beam. These conditions of the analysis enabled the acquisition of approximate compositions of the films at a specified depth from their surface. The microscope operated with a primary electron energy of 5 keV was also used for cross-sectional imaging of the as-deposited and annealed films.

The structure of the films on the Si(100) substrates was characterized by X-ray diffraction (XRD) using a PANalytical X´Pert PRO diffractometer working with a CuK$_\alpha$ radiation ($\lambda$=0.154187 nm) in the Bragg-Brentano geometry.

The hardness and the effective Young´s modulus of the films on the Si(100) substrates were determined from the load vs. displacement curves measured by a Fischerscope H100 microhardness tester with the Vickers diamond indenter at a load of 10 mN. The ratio of the penetration depth to the film thickness was kept below 0.1 for all indentations to eliminate the effect of the substrate.

The electrical resistivity of the films on the pre-oxidized Si(100) substrates was measured by the standard four-point method with a Keithley system with a spacing between the used head's tips of 1.047 mm. Each sample was measured twice, with the specimen rotated 90° before the second measurement.

The extinction coefficient of the films on the Si(100) substrates was determined by a J.A. Woollam Co. variable angle spectroscopic ellipsometer in the wavelength range from 300 to 2000 nm using an incidence angle of 65°, 70° and 75° in reflection. The optical data were then fitted using the WVASE software.

## 3. Results and discussion

In the following sections, a comprehensive analysis of Zr-rich, W/Zr-balanced and W-rich W-Zr and W-Zr-O films is presented. First, the as-deposited structure and cross-sectional microstructure of the W-Zr films (Fig. 1) are introduced. Thereafter, the focus shifts to the oxidation behavior, as examined through thermogravimetric measurements (Fig. 2). The effect of oxidation temperature on the incorporation of oxygen into the W-Zr films is then analyzed (Figs. 3 and 4) along with changes in the structure and cross-sectional microstructure of the resulting W-Zr-O films (Figs. 5 and 6). Finally, the optical extinction coefficient with the electrical resistivity (Figs. 7 and 8) and mechanical properties (Fig. 9) of the W-Zr-O films are presented and discussed.

*3.1. As-deposited (micro)structure*

Fig. 1 shows XRD patterns and SEM cross-sectional micrographs of the as-deposited W-rich, W/Zr balanced, and Zr-rich W-Zr films with decreasing [W]/[Zr] ratio, where [W] and [Zr] stand for the atomic W and Zr contents, respectively. The XRD patterns confirm an amorphous structure in all films as evidenced by a broad, low-intensity peak (amorphous hump). This peak shifts to lower diffraction angles, as the [W]/[Zr] ratio decreases, due to the substitution of smaller W atoms by larger Zr atoms, which results in a change of nearest-neighbor distance of atoms in the amorphous network. The cross-sectional SEM images reveal that all films possess a dense microstructure. The fracture surfaces of the films, obtained under ambient conditions, display vein-like patterns and/or striations to varying degrees, which confirms the metallic glass character of the films.

*3.2 Oxidation behavior*

Thermogravimetric curves of the three W-Zr films annealed in air to 600°C are presented in Fig. 2. It can be seen that a decrease in the [W]/[Zr] ratio results in two notable trends: (i) an earlier onset of oxidation (350°C for the Zr-rich film compared to 400°C for the W-rich film), and (ii) a higher mass gain at the final temperature of 600°C (0.40 mg/cm$^2$ for the Zr-rich film compared to 0.18 mg/cm$^2$ for the W-rich film). These observations are consistent with the higher oxidation tendency of Zr relative to W, as indicated by the formation enthalpies per mole of O atoms for their stoichiometric and substoichiometric oxides (see the values in Table 1).

As we showed in [20], no protective oxide layer forms on the surface of these films upon annealing in air to 600°C and the films possess a dense featureless microstructure in cross sections. Based on these observations, we carried out annealing of the W-rich, W/Zr balanced and Zr-rich films in air to various temperatures from 300°C to 600°C with an increment of 50°C and investigated the evolution of the composition, structure and properties in more detail.

As oxygen diffuses into the films from the ambient synthetic air, its incorporation into the structure occurs through bonding to both Zr and W atoms and results in a gradual mass gain with increasing annealing temperature. Given that the molar volumes (at one metal atom in all structural units) of stoichiometric $ZrO_2$ and $WO_3$ are significantly larger than those of metallic Zr and W ($V_{ZrO2}$ = 21.69 cm$^3$/mol vs. $V_{Zr}$ = 14.02 cm$^3$/mol and $V_{WO3}$ = 32.38 cm$^3$/mol vs. $V_W$ = 9.53 cm$^3$/mol), it is expected that increasing O incorporation will also lead to an increase of the film thickness. This behavior is demonstrated in Fig. 3 for all three films. The thickness expansion follows a trend similar to the mass gain observed in Fig. 2, reaching, for example, up to 58% after annealing in air to 600°C for the Zr-rich film.

Employing EDS on the film surfaces, we quantified the O content, [O], in the top part of the W-Zr-O films after their annealing to the given temperatures (note that the EDS interaction volume for oxygen detection reaches a depth of about 200 nm). Fig. 4 shows the measured [O] in terms of its ratio to that which would correspond to complete oxidation into a mixture of stoichiometric oxides ($WO_3+ZrO_2$). We refer to it here as *O saturation*. The figure is complementary to the whole depth profiles of [O], which are shown in Fig. 5. It can be seen that the film surfaces are not fully saturated by oxygen even after annealing in air to 600 °C (see the discussion below). Furthermore, Fig. 4 reveals the composition-dependent differences. The highest O saturation (69.6% after annealing to 600°C) was achieved for the surface of the Zr-rich film. This is consistent not only with the highest overall mass change (Fig. 2), but also with the fact that a higher Zr content leads to easier saturation (less required O atoms per unit thickness) due to (i) 1.5× lower atomic density of pure Zr compared to pure W, and (ii) 1.5× less O atoms required for oxidation ($ZrO_2$ instead of $WO_3$) at a given number of metal atoms.

Fig. 5 shows SEM cross-sectional views of the W-Zr-O films after annealing in air to 350°C (onset of the mass gain for the Zr-rich film, see Fig. 2), 550°C and 600°C, along with EDS profiles of [O] measured across the film depths. There are two phenomena to point out.

First, independently of the [W]/[Zr] ratio, there is no protective stoichiometric oxide layer on the surface of the films formed. Instead, all surface [O] values are lower than those given by the corresponding weighted average of $WO_3$+$ZrO_2$ (see also the previous Fig. 4), and some oxidation can be observed all the way to the substrate for all films. This can be explained by formation enthalpies, $H_f$ [21,22], given in Table 1. On the one hand, $|H_f|$ of $WO_x$ and $ZrO_x$ per mole of metal atoms is the highest for $WO_3$ and $ZrO_2$, confirming that these stoichiometric phases are prone to form under O-rich conditions. On the other hand, $|H_f|$ of $WO_x$ and $ZrO_x$ per mole of O atoms increases with decreasing $x$ (except extremely O-poor $Zr_6O$), indicating that a homogeneous substoichiometric oxide is thermodynamically preferred to form under the present metal-rich conditions. Note that the presently observed agreement of this thermodynamic preference with our results is far from guaranteed: see, e.g., the formation of a stoichiometric $ZrO_2$ surface layer during the oxidation of pure Zr in [23] and Refs. therein. A case can be made that this captures the fundamental difference between the oxidation of an amorphous and a crystalline material. The oxidation of amorphous W-Zr TFMGs takes place all along its area, making the homogeneous O distribution not only thermodynamically preferred but also relatively kinetically easy. The oxidation of polycrystalline Zr [23] strongly depends on the diffusion of O atoms along the crystal boundaries, making $ZrO_2$ thermodynamically metastable but kinetically prone to form due to the concentration of most O atoms into specific zones.

Second, however, the inevitable combination of the thermodynamic preference to form homogeneous substoichiometric oxides with the finite atomic mobility did not always lead to a depth-independent [O] profile. Instead, there are important differences between individual films, especially at the lowest oxidation temperature of 350 °C, where kinetics is most important. There is a steep depth dependence of [O] in the W-rich film (≈60 at.% on the surface but ≈10 at.% below a depth of ≈600 nm), weaker depth dependence of [O] in the W/Zr-balanced film (≈60 at.% on the surface but ≈18 at.% below a depth of ≈500 nm) and the weakest depth dependence of [O] in the Zr-rich film (≈40 at.% on the surface but ≈22 at.% near the substrate). This can be at least partially explained by the exceptional driving force to form substoichiometric $Zr_3O$ with 25 at.% O (close to the aforementioned 22 at.% near the substrate of the Zr-rich film) under metal-rich conditions. Again, see $H_f$ in Table 1: there is a dramatic difference between $ZrO_2$ (-486 kJ per mole of O atoms) and $Zr_3O$ (-667 kJ per mole of O atoms), while the available $H_f$ differences for $WO_x$ are an order of magnitude smaller.

Increasing annealing temperature leads to increasingly depth-independent substoichiometric [O] regardless of the [W]/[Zr] ratio. As shown in Fig. 5 for 550 °C to 600 °C, the W-Zr-O films exhibit a nearly uniform oxygen distribution across the film thickness. Furthermore, the microstructure of all three W-Zr-O films, irrespective of annealing temperature, remains dense and compact, but with no evidence of features in cross-sectional views characteristic of TFMGs, as shown in Fig. 1.

Fig. 6 presents XRD patterns of all three as-deposited and annealed films, providing evidence that all films retain amorphous structure with no detectable crystalline phases upon annealing in air up to 600°C. However, thermally activated diffusion processes associated with O incorporation induce reorganization of the amorphous structure with increasing annealing temperature. This structural reorganization is reflected in shifts and changes in the shape of the amorphous humps, which vary depending on the [W]/[Zr] ratio.

For the W-rich film, the amorphous hump shifts gradually to higher diffraction angles with increasing annealing temperature. This behavior can be attributed to the relatively uniform incorporation of smaller O atoms (compared to W and Zr) into the dense W-rich amorphous structure, where diffusion is limited. The slower diffusion is consistent with the exceptionally high strength of covalent component of the covalent/metallic bonding associated with W (as well as by other group VI elements, all with high melting points). This is due to its half-filled valence $d$-orbital, which provides relatively localized states responsible for the covalent bonding (the projection of molecular orbitals in pure W on atomic orbitals is close to $5d^56s^1$).

In contrast, a significantly lower [W] in the Zr-rich film facilitates easier atomic diffusion leading to the segregation of the amorphous structure into (Zr-O)-rich and (W-O)-rich regions. This segregation is evidenced by the evolution of the amorphous hump, which splits into two distinct humps as the annealing temperature increases. After annealing to 600°C, one low-intensity hump aligns with the position of the hump of the W-rich film, while the second one is located near 30° 2θ, which corresponds to the characteristic diffraction region of crystalline $ZrO_2$ phases. The evolution of the amorphous structure of the W/Zr-balanced film represents a transitional case between the behavior W-rich and Zr-rich films.

Based on these findings, the transformation of W-Zr TFMGs into ceramic W-Zr-O films is anticipated to alter their functional properties in dependence on O saturation and the [W]/[Zr] ratio. Therefore, the electrical, optical and mechanical properties are discussed in detail in the following sections.

*3.3 Optical and electrical properties*

The effect of O saturation of the three W-Zr-O films (measured for their 200 nm thick top part, see Fig. 4) on their extinction coefficient, $k_{550}$ (measured at 550 nm and also characterizing their top part), and the electrical resistivity, $\rho$, is shown in Fig. 7. In the as-deposited state, these properties are nearly independent of the film composition, with $\rho \approx 1.4 \times 10^{-4}$ $\Omega \cdot$cm and $k_{550} \approx$ 3.3 for all three films. After the annealing in air, several changes are notable. As O atoms are increasingly incorporated into the amorphous structure with increasing annealing temperature, the films become more optically transparent and less electrically conductive. The decrease in $k_{550}$, observed at energies below the band gap of both $ZrO_2$ and $WO_3$, correlates well with increasing O saturation. The dependencies for the individual [W]/[Zr] ratios almost overlap and follow an exponential decay trend. Regarding $\rho$, the O incorporation causes a modest initial increase of about 10–20% relative to the as-deposited state, which persists up to approximately 30% O saturation. Beyond this point, $\rho$ exhibits a sharp increase.

At a given O saturation (not to be confused with a given annealing temperature), the [W]/[Zr] ratio plays only a subtle role in modulating these properties. If anything, the W-rich films exhibit slightly higher $k_{550}$ and $\rho$. The higher $k_{550}$ can be attributed to the narrower energy band gap of $WO_3$ ($\approx 3.15$ eV [24]) compared to $ZrO_2$ ($\approx 5.3$ eV [25]), which is believed to qualitatively hold true also for the substoichiometric oxides of W and Zr. The higher $\rho$ of the W-rich films at intermediate O saturations is likely due to less uniform O depth distribution, with an elevated [O] near the surface (Fig. 5), leading to the formation of a more insulating surface. This effect diminishes at higher levels of O saturation as the O distribution becomes more uniform throughout the film volume.

Fig. 8 provides a direct relationship between $k_{550}$ and $\rho$. It is evident that the data aligns with the previously mentioned trend associated with the different energy band gaps of $WO_3$ and $ZrO_2$ and the corresponding suboxides. Moreover, the figure highlights the versatility of the W-Zr-O films, which can be tailored to achieve desirable combinations of $k_{550}$, ranging from 0.28 to 1.06, and ρ, ranging from 1.7 to 95.7 × $10^{-4}$ Ω·cm.

Using the measured spectral extinction coefficients and refractive indices of the W-Zr-O films in the whole visible range (not shown here), we also calculated the integral luminous transmittance, $T_{lum}$, which quantifies the fraction of visible light transmitted through the films weighted by the human eye's sensitivity [26]. On the one hand, the achievable combinations of electrical and optical properties of the W-Zr-O films cannot compete with the corresponding characteristics of transparent conducting oxide (TCO) films (ρ within $10^{-4}$ Ω·cm and $k_{550}$ <0.001 [27]). On the other hand, these property combinations of the W-Zr-O films are comparable to characteristics of other films studied for deeper reasons than the mere room-temperature conductivity and transmittance, such as magnetron-sputtered $VO_2$ films with the thermochromic semiconductor-to-metal transition [26], see the comparison in Table 2 (for the thickness of 77 nm in the case of $T_{lum}$, not to be confused with the experimental thickness of ≈ 2000 nm).

Let us note that further refinement of the [W]/[Zr] ratio, O saturation and film thickness may enable even more optimized combinations of optical and electrical properties of the W-Zr-O films.

*3.4 Mechanical properties*

The mechanical properties of the W-Zr-O films, which are a critical factor for the wear resistance, were evaluated from indentation measurements. Figs. 9a and 9b illustrate the hardness, $H$, and effective Young's modulus, $E^*$, as a function of O saturation. Unlike the nearly [W]/[Zr]-independent electrical and optical properties, one can notice significant dependences on the [W]/[Zr] ratio in this case. In the as-deposited state, $H$ values are 11.2 GPa, 9.8 GPa and 7.1 GPa, while $E^*$ values are 125 GPa, 112 GPa and 93 GPa for the W-rich, W/Zr-balanced and Zr-rich films, respectively. These differences can be explained by the more covalent character of the covalent/metallic bonds associated with W, as discussed in Sec. 3.2.

As O saturation increases, both mechanical properties show similar evolutionary trends, still reflecting the differences in the as-deposited state. The initial O incorporation enhances both $H$ and $E^*$, which are, however, maintained relatively constant up to approximately 30% O saturation. The relatively highest enhancement, with a 42% increase in $H$ and a 16% increase in $E^*$, is observed for the Zr-rich films. This improvement is likely linked to the segregation of the amorphous structure into more abundant (Zr-O)-rich regions with higher $H$ compared to (W-O)-rich regions (Fig. 6). Note that $H$ of magnetron-sputtered $ZrO_2$ films ranges from 12–17 GPa [28], while that of $WO_3$ films varies between 7–15 GPa [29,30].

When O saturation increases beyond 30%, both mechanical properties rise sharply to maximum values, which are achieved at approximately 50% O saturation. Hence, most of the transition from the covalent/metallic state to a largely ionically bonded state arguably takes place in this compositional range (let us also recall the sharply increasing $\rho$ above exactly the same threshold of 30%). The maximum values are remarkably consistent for all three [W]/[Zr] ratios: $H$ reaches approximately 17.5 GPa, while $E^*$ is around 170 GPa. Interestingly, a slight decline of both mechanical properties is observed with a further increase in O saturation beyond 50%. This reveals that the maximum values of the mechanical properties in the W-Zr-O system might be achieved for an optimized substoichiometric composition.

The W-Zr-O films demonstrate mechanical properties superior to those of other electrically conductive and optically transparent films. With $H$ ranging from 16.0 to 17.5 GPa, they surpass thermochromic thermochromic $VO_2$ films (12 GPa [31]), Al-doped ZnO (AZO) films (8–11 GPa [32,33]), Sn-doped $In_2O_3$ (ITO) films (6.5 GPa [34], 16 GPa [35]) and InGaZnO (IGZO) films (3 GPa [36]).

In addition, the W-Zr-O films exhibit noteworthy values of the $H/E^*$ ratio, as illustrated in Fig. 9c. The incorporation of oxygen into the films enhances this ratio above 0.1, especially in the 30–50% oxygen saturation range. These elevated values imply their greater capacity to undergo elastic strain before fracturing, which is a characteristic beneficial in applications requiring both flexibility and durability.

## 4. Conclusions

A two-step process, comprising magnetron-sputter deposition of W-Zr TFMGs followed by controlled oxidation, was successfully employed to prepare ceramic W-Zr-O films. In this study, the oxidation behavior of three W-Zr TFMGs with varying Zr contents (32, 48, and 61 at.%) was systematically investigated during annealing in synthetic air to various temperatures up to 600°C. Particular attention was paid to understanding oxidation-induced phenomena, especially those associated with changes in the composition and structure. The resulting effects on the optical, electrical, and mechanical properties of the films were clearly identified and thoroughly analyzed.

The main results demonstrate that oxidation process leads to the transformation of W-Zr TFMGs into W-Zr-O films with substoichiometric composition and dense and compact amorphous structures. This transformation is governed by thermodynamic factors, as the more negative formation enthalpies of substoichiometric $ZrO_x$ and $WO_x$ hinder the formation of stoichiometric surface oxide layers that could act as passivating diffusion barriers. The O content and its distribution across the film thickness can be finely controlled by the annealing temperature and [W]/[Zr] content ratio in the W-Zr TFMGs. A higher Zr content promotes O diffusion into the amorphous structure and enhances its depth uniformity.

The optical and electrical properties are strongly influenced by O saturation within the amorphous structure and not so much by the [W]/[Zr] ratio, while the mechanical properties also depend on this ratio. As O saturation increases, the films become more optically transparent and less electrically conductive. The relationship between the extinction coefficient and the electrical resistivity highlights the versatility of the W-Zr-O films, enabling their tailoring for desirable combinations of optical transparency (extinction coefficient: 0.28–1.06) and electrical conductivity (resistivity: $1.7$–$95.7 \times 10^{-4}$ $\Omega \cdot$cm). The films also exhibit remarkable hardness values ranging from 16.0 to 17.5 GPa, which surpass those of most electrically conductive and optically transparent coatings. Interestingly, the hardness reaches maximum value at approximately 50% O saturation independently of the [W]/[Zr] ratio. For higher O saturations, the hardness slightly decreases.

These exceptional properties demonstrate the potential of W-Zr-O films in advanced optoelectronic and protective applications. Their ability to combine the optical transparency and the electrical resistivity in a promising way and their superior hardness makes them suitable for use in devices requiring durable coatings with multifunctional capabilities.

**Acknowledgments**

This work was supported by the Czech Science Foundation under Project No. GA22-18760S and by the project QM4ST under Project No. CZ.02.01.01/00/22_008/0004572 funded by Program Johannes Amos Comenius, call Excellent Research. The authors also acknowledge Dr. Z. Soukup and D. Thakur for their assistance with sample preparation.

# Tables

**Table 1.** Formation enthalpy, $H_f$, of $WO_x$ and $ZrO_x$ phases per mole of atoms, per mole of metal (M) atoms and per mole of O atoms. The data are calculated from experimental enthalpies per mole of structural units ($WO_x$ [21]) and ab-initio enthalpies per mole of atoms ($ZrO_x$ [22]).

|  | $H_f$ (kJ/mol) | | |
| --- | --- | --- | --- |
|  | per mole of atoms | per mole of M atoms | per mole of O atoms |
| $WO_x$ (exp.) [21] | | | |
| $WO_3$ | -211 | -843 | -281 |
| $WO_2$ | -197 | -590 | -295 |
| $ZrO_x$ (calc.) [22] | | | |
| $ZrO_2$ | -324 | -972 | -486 |
| $ZrO$ | -252 | -504 | -504 |
| $Zr_2O$ | -172 | -258 | -515 |
| $Zr_3O$ | -167 | -223 | -667 |
| $Zr_6O$ | -76 | -89 | -533 |

**Table 2.** Electrical resistivity, $\rho$, extinction coefficient, $k_{550}$, and integral luminous transmittance, $T_{lum}$, of selected W-rich, W/Zr balanced and Zr-rich W-Zr-O films with different O saturations, along with thermochromic $VO_2$ films below (@ 25 °C) and above (@ 100 °C) the semiconductor-to-metal transition temperature. The experimental values of $\rho$ and $k_{550}$ for W-Zr-O films correspond to a film thickness ≈ 2000 nm, whereas the values of $T_{lum}$ were calculated for 77 nm thick films to align with the experimentally measured values for $VO_2$ films.

|  | O saturation (%) | $\rho$ ($\Omega \cdot cm$) | $k_{550}$ (-) | $T_{lum}$ (%) |
| --- | --- | --- | --- | --- |
| $VO_2$ [26] | | | | |
| @ 25 °C | 80.0 | $9.3 \times 10^{-3}$ | 0.48 | 30.0 |
| @ 100 °C | 80.0 | $4.1 \times 10^{-4}$ | 0.61 | 27.0 |
| W-Zr-O | | | | |
| W-rich | 38.6 | $3.7 \times 10^{-4}$ | 1.06 | 14.3 |
| W/Zr balanced | 47.5 | $1.1 \times 10^{-3}$ | 0.67 | 26.1 |
| Zr-rich | 47.4 | $2.4 \times 10^{-4}$ | 0.62 | 26.5 |
| Zr-rich | 51.9 | $5.4 \times 10^{-4}$ | 0.56 | 29.5 |
| Zr-rich | 58.1 | $2.8 \times 10^{-3}$ | 0.29 | 49.2 |

# Figure captions

**Fig.1.** XRD patterns and SEM cross-sectional micrographs of as-deposited W-rich, W/Zr-balanced and Zr-rich W-Zr thin-film metallic glasses.

**Fig. 2.** Thermogravimetric oxidation curves of W-Zr thin-film metallic glasses measured in air up to 600°C. The dotted lines denote the annealing temperatures selected for subsequent analyses.

**Fig. 3.** Thickness change caused by oxidation of W-rich, W/Zr-balanced and Zr-rich W-Zr thin-film metallic glasses during post-deposition annealing in air to various temperatures. The thickness change is related to the as-deposited film thickness.

**Fig. 4.** Oxygen saturation in W-rich, W/Zr-balanced and Zr-rich W-Zr-O films resulting from post-deposition annealing of W-Zr thin-film metallic glasses in air to various temperatures. A saturation level of 100% corresponds to the oxygen content in the films (measured from the film surfaces by EDS) associated with their complete oxidation into a mixture of stoichiometric oxides ($WO_3+ZrO_2$) at the given W and Zr contents. Note that the EDS interaction volume for oxygen detection reaches a depth of about 200 nm.

**Fig. 5.** Representative examples of the cross-sectional microstructure of W-rich, W/Zr-balanced and Zr-rich W-Zr-O films and EDS depth profiles of their oxygen contents after post-deposition annealing in air to various temperatures.

**Fig. 6.** XRD patterns of W-rich, W/Zr-balanced and Zr-rich W-Zr-O films formed during post-deposition annealing to various temperatures. The dashed line denotes the position of the maximum amorphous hump of the as-deposited W-rich film.

**Fig. 7.** (a) Extinction coefficient and (b) electrical resistivity of W-rich, W/Zr-balanced and Zr-rich W-Zr-O films plotted as a function of oxygen saturation. The extinction coefficient was measured at 550 nm and characterizes the top part of the films.

**Fig. 8.** Relationships between the electrical resistivity and the extinction coefficient of W-rich, W/Zr-balanced and Zr-rich W-Zr-O films. The extinction coefficient was measured at 550 nm and characterizes the top part of the films.

**Fig. 9.** (a) Hardness, (b) effective Young's modulus and (c) their ratio for W-rich, W/Zr-balanced and Zr-rich W-Zr-O films plotted as a function of oxygen saturation.

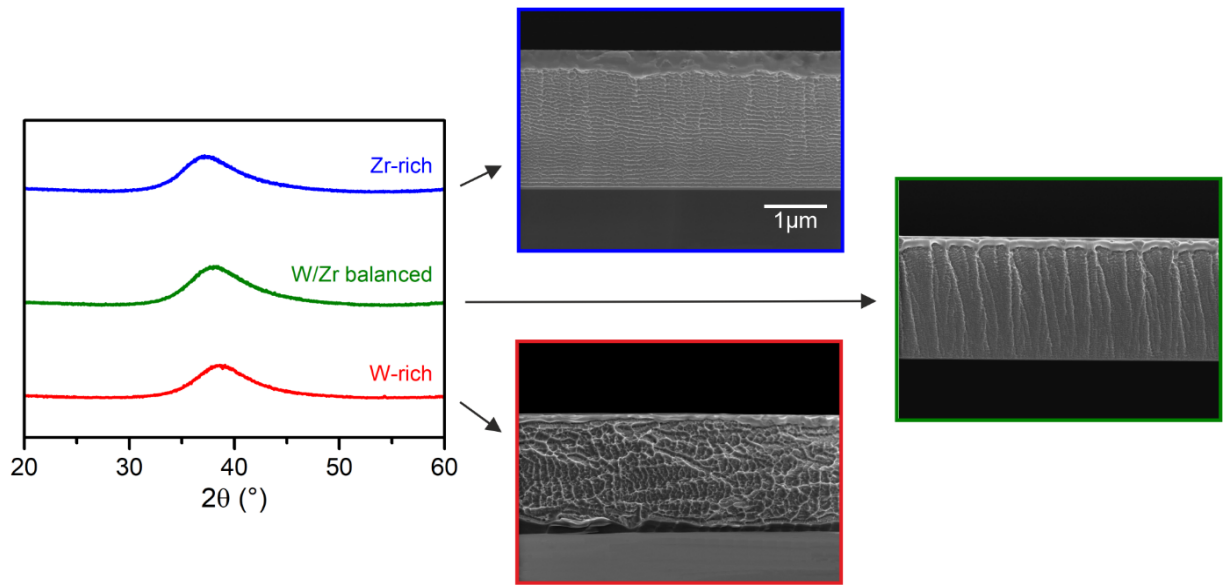

**Fig.1.** XRD patterns and SEM cross-sectional micrographs of as-deposited W-rich, W/Zr-balanced and Zr-rich W-Zr thin-film metallic glasses.

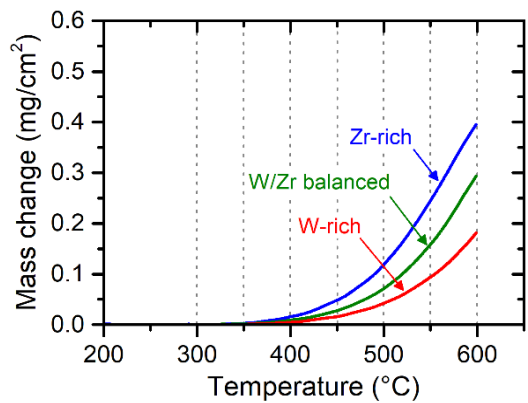

**Fig. 2.** Thermogravimetric oxidation curves of W-Zr thin-film metallic glasses measured in air up to 600°C. The dotted lines denote the annealing temperatures selected for subsequent analyses.

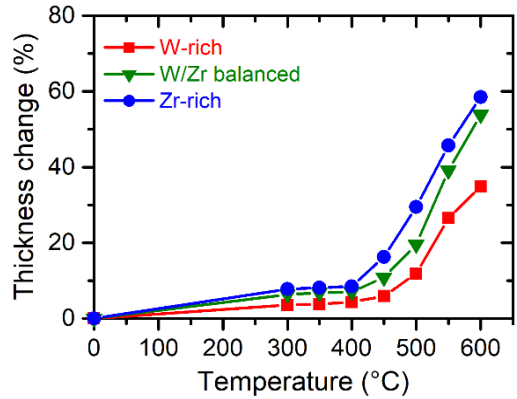

**Fig. 3.** Thickness change caused by oxidation of W-rich, W/Zr-balanced and Zr-rich W-Zr thin-film metallic glasses during post-deposition annealing in air to various temperatures. The thickness change is related to the as-deposited film thickness.

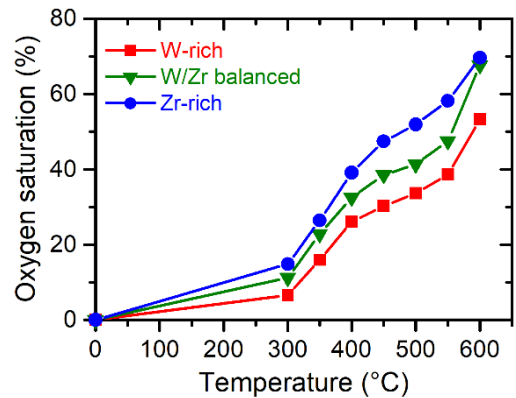

**Fig. 4.** Oxygen saturation in W-rich, W/Zr-balanced and Zr-rich W-Zr-O films resulting from post-deposition annealing of W-Zr thin-film metallic glasses in air to various temperatures. A saturation level of 100% corresponds to the oxygen content in the films (measured from the film surfaces by EDS) associated with their complete oxidation into a mixture of stoichiometric oxides ($WO_3$+$ZrO_2$) at the given W and Zr contents. Note that the EDS interaction volume for oxygen detection reaches a depth of about 200 nm.

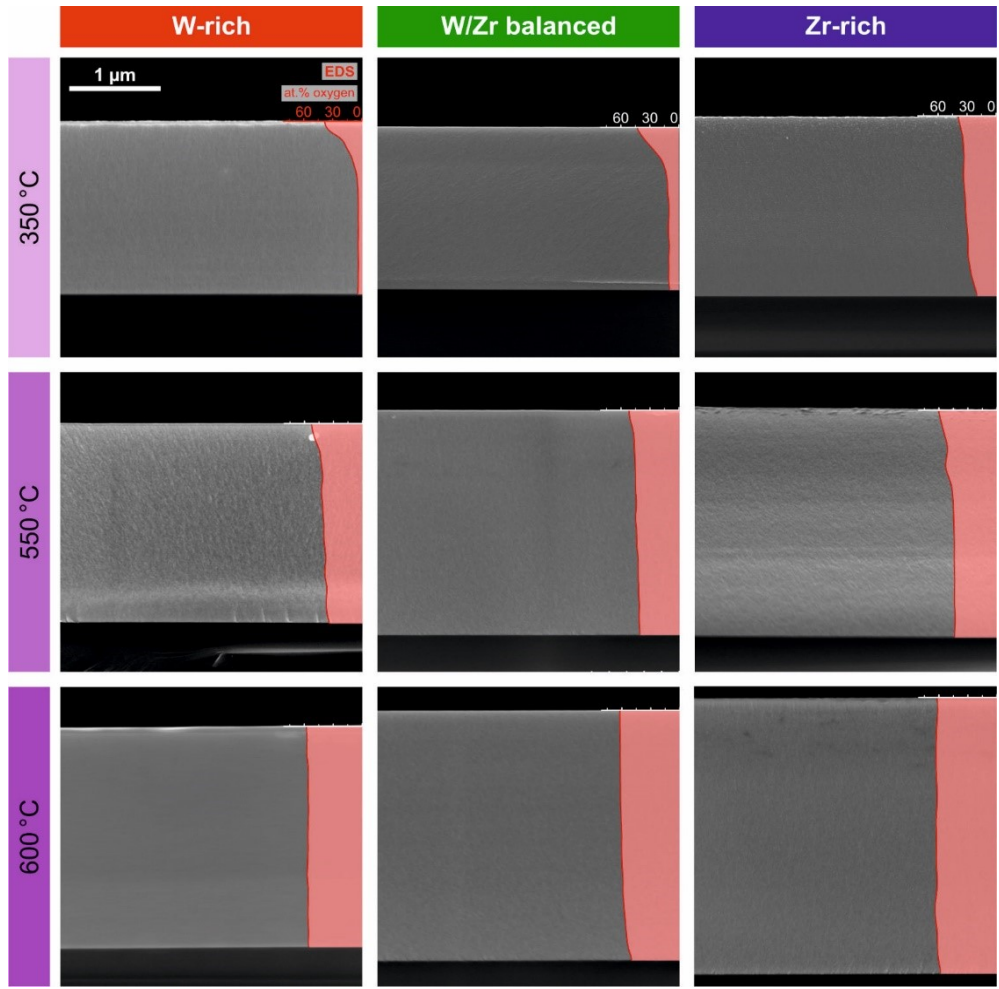

**Fig. 5.** Representative examples of the cross-sectional microstructure of W-rich, W/Zr-balanced and Zr-rich W-Zr-O films and EDS depth profiles of their oxygen contents after post-deposition annealing in air to various temperatures.

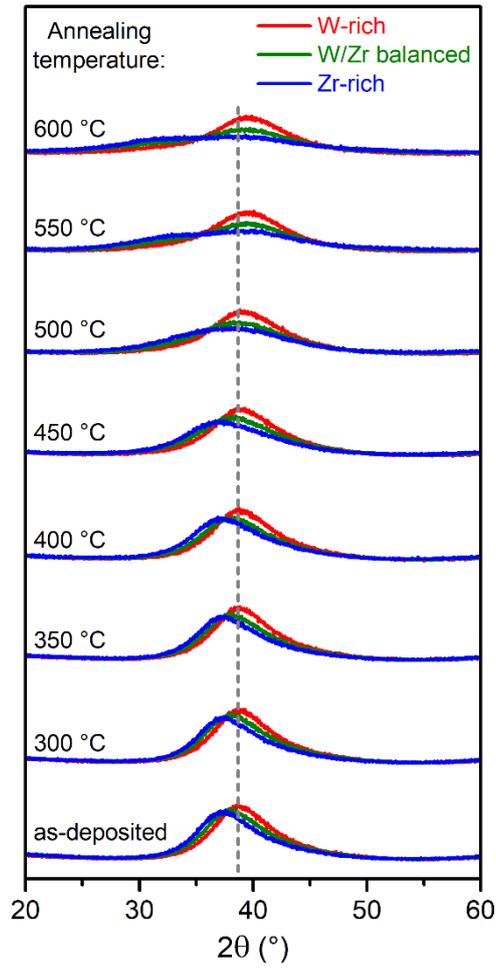

**Fig. 6.** XRD patterns of W-rich, W/Zr-balanced and Zr-rich W-Zr-O films formed during post-deposition annealing to various temperatures. The dashed line denotes the position of the maximum amorphous hump of the as-deposited W-rich film.

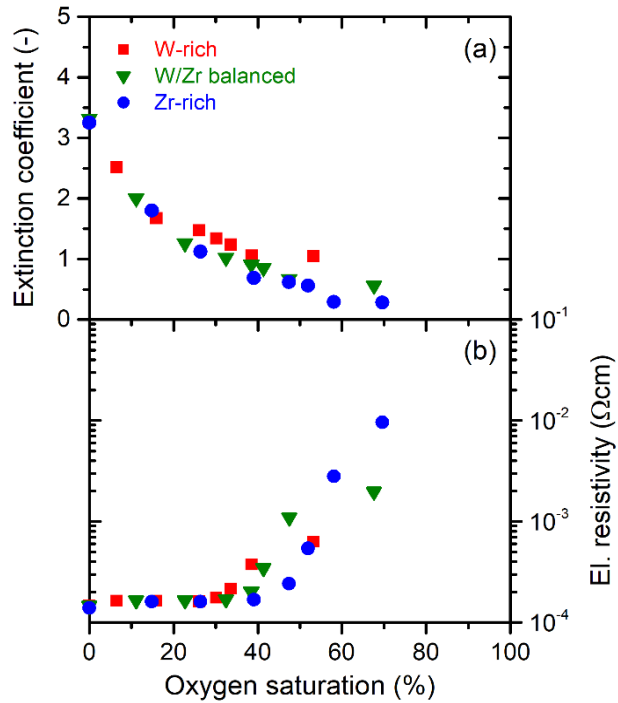

**Fig. 7.** (a) Extinction coefficient and (b) electrical resistivity of W-rich, W/Zr-balanced and Zr-rich W-Zr-O films plotted as a function of oxygen saturation. The extinction coefficient was measured at 550 nm and characterizes the top part of the films.

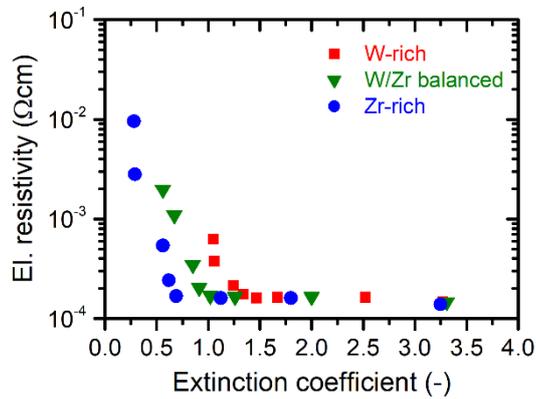

**Fig. 8.** Relationships between the electrical resistivity and the extinction coefficient of W-rich, W/Zr-balanced and Zr-rich W-Zr-O films. The extinction coefficient was measured at 550 nm and characterizes the top part of the films.

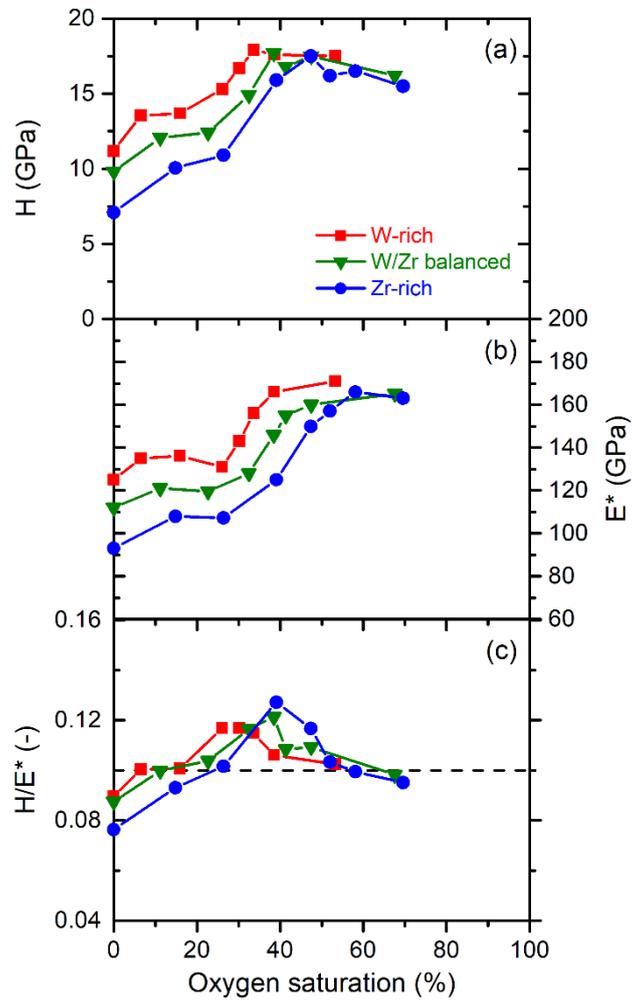

**Fig. 9.** (a) Hardness, (b) effective Young's modulus and (c) their ratio for W-rich, W/Zr-balanced and Zr-rich W-Zr-O films plotted as a function of oxygen saturation.